\documentclass[epj]{svjour}

\usepackage{bm}
\usepackage{amssymb}
\usepackage{color}
\usepackage{mathtools}
\usepackage{tikz}
\usepackage{url}
\usepackage{hyperref}
\usepackage{cite}
\usepackage{xcolor}
\usepackage{tikz}
\usepackage{amsmath}
\usepackage{float}
\usepackage{diagbox}
\usepackage{pgfplots}
\usepackage{subcaption}

\usepackage{tikz}
\usetikzlibrary{decorations.pathmorphing,decorations.markings}
\usetikzlibrary{decorations.pathreplacing}
\usetikzlibrary{positioning, calc, shapes}
\tikzset{scalar/.style={draw=darkred,dashed, dash pattern=on 3pt off 3pt},
    vectorb/.style={decorate, decoration={snake,amplitude=1pt, segment
    length=5pt}, draw=darkgreen}, particle/.style={draw=black,
    postaction={decorate}, decoration={markings,mark=at position 0.5 with
    {\arrow[draw=black]{>}}}}, antiparticle/.style={draw=black,
    postaction={decorate}, decoration={markings,mark=at position 0.5 with
    {\arrow[draw=black]{<}}}}, gluon/.style={decorate, draw=darkblue,
    decoration={coil,amplitude=2.7pt, segment length=3.2pt}},
    gluonflip/.style={decorate, draw=darkblue,
    decoration={coil,amplitude=-2.7pt, segment length=3.2pt}},
    gluonout/.style={decorate, draw=blue, decoration={coil,amplitude=4pt,
    segment length=5pt}},}
\usetikzlibrary{shapes.geometric, arrows}
    \tikzstyle{startstop} = [rectangle, rounded corners, minimum width=5cm,text
    centered, draw=black, fill=red!30] \tikzstyle{roundbox} = [rectangle,
    rounded corners, minimum height=0.6cm, minimum width=5cm,text centered,
    draw=black, fill=blue!30] \tikzstyle{check} = [rectangle, minimum
    width=3.8cm, text centered, draw=black, fill=blue!30] \tikzstyle{iout} =
    [trapezium, trapezium left angle=-70, trapezium right angle=-110, minimum
    width=3cm, text centered, draw=black, fill=blue!30] \tikzstyle{io} =
    [trapezium, trapezium left angle=70, trapezium right angle=110, minimum
    width=3cm, minimum height=1cm, text centered, draw=black, fill=blue!30]
    \tikzstyle{nnpdf} = [rectangle, minimum width=3cm, minimum height=1cm, text
    centered, draw=black, fill=orange!30] \tikzstyle{n3py} = [rectangle, minimum
    width=3cm, minimum height=1cm, text centered, draw=black, fill=green!30]
    \tikzstyle{n3cpp} = [rectangle, minimum width=3cm, minimum height=1cm, text
    centered, draw=black, fill=blue!30] \tikzstyle{procblue} = [rectangle,
    minimum width=3cm, minimum height=1cm, text centered, draw=black,
    fill=blue!30] \tikzstyle{fitted} = [rectangle, minimum width=5cm, minimum
    height=1cm, text centered, draw=black, fill=red!30] \tikzstyle{fixed} =
    [rectangle, minimum width=5cm, minimum height=1cm, text centered,
    draw=black, fill=blue!30] \tikzstyle{arrow} = [thick,->,>=stealth]
    \tikzstyle{decision} = [diamond, draw, text badly centered, inner sep=3pt]
    \tikzstyle{tbox} = [draw, rounded corners] \tikzstyle{operations} =
    [rectangle, rounded corners, minimum width=2cm,text centered, draw=black,
    fill=red!30] \tikzstyle{roundtext} = [rectangle, rounded corners, minimum
    width=2cm, minimum height=0.8cm, text centered, draw=black, fill=red!30]

\begin{document}

\title{{\tt MadFlow}: automating Monte Carlo simulation on GPU for particle
    physics processes}

\author{Stefano Carrazza\inst{1,2,3}, Juan Cruz-Martinez\inst{1},  Marco
    Rossi\inst{1,2}, Marco Zaro\inst{1}}
%
%
\institute{TIF Lab, Dipartimento di Fisica, Universit\`a degli Studi di Milano
    and INFN Sezione di Milano. \and CERN Theoretical Physics Department and
    OpenLab, CH-1211 Geneva 23, Switzerland. \and Quantum Research Centre,
    Technology Innovation Institute, Abu Dhabi, UAE.}

\date{Received: date / Revised version: date}
\abstract{We present {\tt MadFlow}, a first general multi-purpose framework for
    Monte Carlo (MC) event simulation of particle physics processes designed to
    take full advantage of hardware accelerators, in particular, graphics
    processing units (GPUs).
    The automation process of generating all the required components for MC
    simulation of a generic physics process and its deployment on hardware
    accelerator is still a big challenge nowadays.
    In order to solve this challenge, we design a workflow and code library
    which provides to the user the possibility to simulate custom processes
    through the MadGraph5\_aMC@NLO framework and a plugin for the generation and
    exporting of specialized code in a GPU-like format. The exported code
    includes analytic expressions for matrix elements and phase space.
    The simulation is performed using the VegasFlow and PDFFlow libraries which
    deploy automatically the full simulation on systems with different hardware
    acceleration capabilities, such as multi-threading CPU, single-GPU and
    multi-GPU setups.
    The package also provides an asynchronous unweighted events procedure to
    store simulation results.
    Crucially, although only Leading Order is automatized, the library
    provides all ingredients necessary to build full complex Monte Carlo simulators
    in a modern, extensible and maintainable way.
    We show simulation results at leading-order for multiple processes on
    different hardware configurations.
} 

\authorrunning{S.C, J.C.M, M.R., M.Z.}

\maketitle
\section{Introduction}

The popularity of hardware accelerators, such as graphics processing units
(GPU), has quickly increased in the last years thanks to the exceptional
performance benefits and efficiency achieved in scientific and industrial
applications. Furthermore, new code frameworks based on hardware accelerators
have been designed in order to simplify the implementation of algorithms and
models in particular in the context of Artificial Intelligence applications.

If we consider the research domain of High Energy Physics, we can observe
several examples of applications that could benefit from the conversion or
systematic implementation of existing algorithms and code libraries on GPU. Some
examples have already been successfully published, such as deep learning
applications in experimental physics~\cite{Albertsson:2018maf}, where
astonishing performance improvements were obtained thanks to the employment of
GPUs.

Despite the HEP community interest in providing computational tools for
experimental setups, we still observe a growing trend towards the increase of
computational time required to solve complex
problems~\cite{Niehues:2018was,Hoche:2013zja} in particular for Monte Carlo (MC)
simulation of particle physics processes.
Moreover, this growing trend is further increased by the current state of the
art implementations of MC simulation libraries, which still today rely
almost exclusively, on CPU architecture~\cite{Gleisberg:2008ta, Alwall:2014hca,
    Frederix:2018nkq, Campbell:2019dru}.
This is despite the fact that the parallel nature of MC simulations makes
them the perfect target for hardware accelerators.
Attempts at porting MC simulation libraries to GPUs have shown
quite promising results, but they have been limited
in scope~\cite{Hagiwara:2009aq,Hagiwara:2009cy,Hagiwara:2013oka, Grasseau:2019sih, Bothmann:2021nch}.

From a practical perspective, in order to write a competitive GPU-capable full parton-level
MC by any measure with existing tools, there are at least five required
ingredients: (i) an integrator, able to parallelize over the number of events;
(ii) a GPU-capable parton distribution function (PDF) interpolation tool; (iii)
an efficient phase space generator, which should generate valid phase space
points on GPU, apply any fiducial cuts; (iv) finally evaluate the matrix element
squared for the target processes; (v) an efficient asynchronous output storage
system for observables, such as histograms and Les Houches event files.

In recent previous works, we have developed open-source tools that provide the
ground basis for the implementation of an automatic Monte Carlo simulation
framework for HEP addressing some of the aforementioned issues:
VegasFlow~\cite{Carrazza:2020rdn,Carrazza:2020loc} and
PDFFlow~\cite{Carrazza:2020qwu,Rossi:2020sbh}. The first package, VegasFlow, is
a new software for fast evaluation of high dimensional integrals based on Monte
Carlo integration techniques designed for platforms with hardware accelerators.
This project allows developers to delegate all complicated aspects of hardware
or platform implementation to the library, reducing development time due to
maintainability and debugging. On the other hand, PDFFlow is a library which
provides fast evaluation of parton distribution functions (PDFs) designed for
platforms with hardware accelerators following the design idea inspired from
VegasFlow. The availability of both packages completes respectively points (i)
and (ii) above.

The goal of this work is to address and propose an integrated technical solution
for all points above, following the original idea presented
in~\cite{Carrazza:2021zug}. We call {\tt
MadFlow}~\cite{juan_m_cruz_martinez_2021_4954376} the open-source software
library which implements this automatic pipeline for GPU deployment of Monte
Carlo simulation. It combines the matrix elements
expressions generated by the MadGraph5\_aMC@NLO (MG5\_aMC)~\cite{Alwall:2014hca,
Frederix:2018nkq} framework with the VegasFlow and PDFFlow efficient simulation
tool for hardware accelerators. {\tt MadFlow} by design, opens the possibility
to study and benchmark multiple approaches to Monte Carlo integration based on
distributed hardware, and, in future, new algorithms based on deep learning
techniques.

This work is structured as follows. In Section~\ref{sec:methodology} we describe
the technical implementation of {\tt MadFlow}. In Section~\ref{sec:results} we
compare and benchmark results. Finally, in Section~\ref{sec:outlook} we present
our conclusion and future development roadmap.

\section{Implementation}
\label{sec:methodology}

\subsection{The {\tt MadFlow} concept}

{\tt MadFlow} is an open-source python
package~\cite{juan_m_cruz_martinez_2021_4954376}, which provides the user a
simple tool for parton-level Monte Carlo simulation on hardware accelerators.
The original concept of {\tt MadFlow} is to keep usability and maintainability
as simple as possible thanks to the new technologies and frameworks currently
available today.

From the user's point of view, the effort and time required to start using
    {\tt Madflow} and write a fully-working Leading Order Monte Carlo simulation
is limited to the installation of the package and its dependencies.
The library provides a quick-start script which
takes care of the generation of all required modules and code dependencies for
running the simulation on the user's hardware, including multi-threading CPU,
single GPU and multi-GPU setups for both NVIDIA and AMD products.

On the other hand, from the developer perspective, the {\tt MadFlow} code is
based on primitives actively maintained by the community and its design is
modular: all components presented in the next paragraphs can be modified and
extended with minor effort, thanks to an uniform standardization of the
frameworks and documentation.

\subsection{The {\tt MadFlow} code design}

Nowadays many research groups rely on very extensive and complex code bases,
thus learning how to use an equally complicated framework might require time and
expertise which may not be feasible for everyone. For example, consider the
investment in time required for training of new doctoral students or
researchers. Therefore, in order to accelerate the adoption of hardware
accelerators within the hep-ph community we design the {\tt MadFlow} Monte Carlo
implementation on GPU with maintainability and developer-friendliness as a major
target feature.

We consider the MG5\_aMC framework as the entry point of our procedure. MG5\_aMC
is a meta-code written in Python, that generates automatically the code in a
low-level language to be employed for the simulation of arbitrary scattering
processes at colliders, in the Standard Model or beyond. MG5\_aMC relies on
general procedures and methods, without being tailored to a specific process,
class of processes, or physics model. Besides the generation of tree-level
matrix elements, MG5\_aMC gives also the possibility to the user to include
Next-to-Leading order corrections, both due to strong and electroweak
interactions (including matching to parton-shower for the former). However, in
this paper we will limit ourselves to the case of tree-level matrix elements.

The workflow of MG5\_aMC is the following: a model, written in the Universal
Feynrules Output (UFO) format~\cite{Degrande:2011ua}, is loaded, which contains
all the informations on the underlying theory, including the Feynman rules.
Starting from the model, Feynman diagrams are generated, and the corresponding
expressions for the matrix elements are written in process-specific files. The
various parts of the Feynman diagrams (external wavefunctions, vertices,
propagators, etc.) are evaluated via the ALOHA routines~\cite{deAquino:2011ub}
(with the introduction of MadGraph5~\cite{Alwall:2011uj} ALOHA supersedes the
HELAS routines~\cite{Murayama:1992gi}).

It is then natural to consider MG5\_aMC, in particular ALOHA, as a backend of
matrix elements for GPU for a future general purpose parton-level GPU MC
generator. We should note there is an independent effort dedicated to porting
MG5\_aMC to GPU~\cite{madgraph4gpu}. The {\tt MadFlow} project differs from
``Madgraph 4 GPU'' for two main reasons:
the interest in providing a full MC simulator based on modern software
which can automatically be deployed in different hardware configurations
and the need of a technical solution which simplifies
maintainability and does not require specialized GPU knowledge from the
developer and user point of view.
However, we must note CUDA-based libraries are compatible with TensorFlow and thus
it is technically possible to use ``Madgraph 4 GPU'' matrix elements
within {\tt MadFlow}, maximizing the advantages both codes provide.

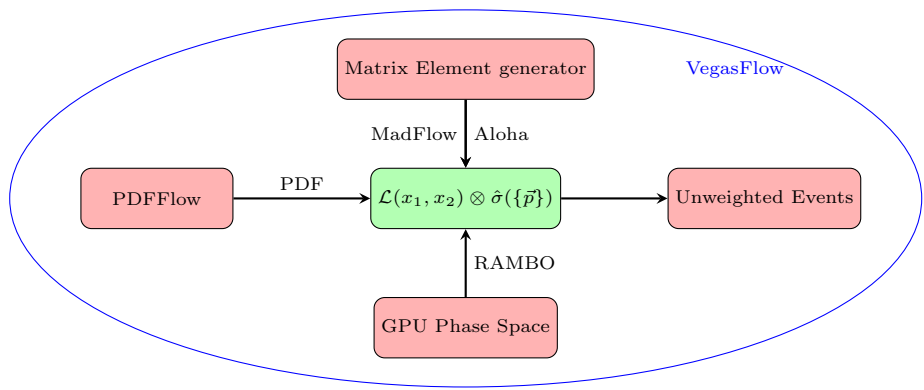
\begin{figure*}
    \begin{center}
        \begin{tikzpicture}\scriptsize \node[roundtext, fill=green!30] (central)
            {$\mathcal{L}(x_{1},x_{2})\otimes \hat{\sigma}(\{\vec{p}\})$};

            \node[roundtext, above = 0.9cm of central] (madgraph) {Matrix
                Element generator}; \node[roundtext, left  = 1.8cm of central]
            (pdfflow) {PDFFlow}; \node[roundtext, right = 1.4cm of central]
            (uwgt) {Unweighted Events}; \node[roundtext, below = 0.9cm of
                central] (ps) {GPU Phase Space};

            \draw[arrow] (madgraph) -- node[right] {Aloha} (central);
            \draw[arrow] (madgraph) -- node[left] {MadFlow} (central);
            \draw[arrow] (pdfflow) -- node[above] {PDF} (central); \draw[arrow]
            (ps) -- node[right] {RAMBO} (central); \draw[arrow] (central) --
            (uwgt);

            \draw[blue] (central) ellipse (6cm and 2.5cm);

            \node[blue, right = 1.1cm of madgraph] (vflow) {VegasFlow};

        \end{tikzpicture}
    \end{center}
    \caption{\label{fig:scheme}Schematic workflow for the implementation of {\tt
                MadFlow}. The software automates the generation of a custom process
        using the standard MG5\_aMC framework API and exports the relevant
        code in a specific format for VegasFlow and PDFFlow integration.}
\end{figure*}

In Figure~\ref{fig:scheme} we show the modules involved in the current implementation of
    {\tt MadFlow}. The process is driven by a {\tt MadFlow} script which
generates a custom process using the MG5\_aMC standard framework API and exports
the relevant code for the analytic matrix elements and phase space expressions
in python, using the syntax defined by VegasFlow. In terms of code
implementation this step has required the development of a MG5\_aMC plugin, which
consists of an exporter module to write the matrix element and the ALOHA
routines in Python, fulfilling the requirements imposed by VegasFlow and PDFFlow
using TensorFlow~\cite{tensorflow2015:whitepaper} primitives. The main
difficulty consists in converting sequential functions into vectorized
functions.
During the {\tt MadFlow} development we have performed several numerical and
performance benchmarks in order to avoid potential bugs.

After the conversion step into the specified format is performed, the exported
python code is incorporated into a VegasFlow device-agnostic integration
algorithm which executes the event generation pipeline from the generation of
random numbers, computation of the phase space kinematics, matrix element
evaluation and histogram accumulation.

\subsection{The evaluation of matrix elements routines}

In {\tt MadFlow}, the matrix elements evaluations follows the original MG5\_aMC
implementation: a {\tt Matrix} class is produced by the Python exporter plugin
module. Its {\tt smatrix} method links together the needed Feynman rules to
compute the requested matrix element: it loops over initial and final state
particle helicities and aggregates their contribution to the final squared
amplitude.

The matrix element vectorization requires to replace the ALOHA waveforms and
vertices routines abiding by the TensorFlow ControlFlow rules. Although this
process is straightforward for vertices Feynman rules, being mostly comprised by
elementary algebraic operations, the implementation of particle waveforms
functions is subject to several conditional control statements that
make the task harder as GPUs suffer considerably from branching.

\subsection{Phase-space generation}
\label{sec:ps}

The integration phase-space is generated using a vectorized implementation of
the RAMBO algorithm~\cite{Kleiss:1985gy} which makes it suitable for
hardware accelerators.

RAMBO maps the random variables of the integrator library to a flat phase-space
with which the {\tt smatrix} method of the matrix element can be evaluated.
While this means {\tt MadFlow} can produce results for any number of particles,
the generated phase space doesn't take into account the topology of the process.
As a result, for a great number of final-state particles the number of events
required to get a reasonably precise result is much larger than what would
be required with other Monte Carlos.

More complex and efficient phase-space generators will be developed for future releases
of {\tt MadFlow}.

\subsection{Unweighted events exporter}
\label{sec:lhe}

The {\tt Madflow} event generator is equipped with a Les Houches Event (LHE)
writer module that provides a class to store events in the LHE 3.0 file
format~\cite{butterworth2014les}. The LHE writer class operates asynchronously
in a separated thread from the VegasFlow integration, thus ensuring that the
integrator computational performance is not harmed by IO limitations.
The module works by collecting all the unweighed events generated by the integrator
and applying an unweighting procedure employing the module provided by MG5\_aMC. The
final result is a (compressed) LHE file, with unweighted events.

We note that in this implementation, however, the unweighting efficiency is rather low
(around $5 \%$) because of the non-optimised phase space which relies on RAMBO.

\subsection{Scope}

The goal of {\tt MadFlow} is to provide the foundation for future high precision
Monte Carlo simulation (of higher orders or otherwise) so they can efficiently take
advantage of hardware developments.

In its current form, {\tt MadFlow} provides the necessary tools for the computation
of Leading Order (LO) calculations fully automatically for any number of
particles.\footnote{The code is still in beta testing and some corner cases may fail due to lack of tests.}
Higher order calculations can be implemented by building upon the provided LO template.
Parameters and model definitions are provided by MG5\_aMC-compatible parameters card
and so the same options and models can be used or defined.

An important caveat must be considered when trying to integrate more complex processes.
The provided phase-space is flat and does not take into account the topology of the diagrams
(see Section~\ref{sec:ps}) and therefore becomes inefficient with large multiplicities.
As a result a very large number of events might be necessary to reduce the Monte Carlo error,
cancelling the benefits of running on a GPU.
Therefore, for processes with many particles in the final state
or when using the tree level amplitudes for Next-to-Leading Order (NLO) calculations
writing an specialized phase-space is recommended~\cite{Kleiss:1994qy}.
For some recent developments on this matter, see also~\cite{Ostrolenk:2021tfz}.

In summary, while due to some of its limitations {\tt MadFlow} cannot provide yet
efficient calculations, it can quickstart the development of efficient
fixed-order Monte Carlo simulators in hardware accelerators.

\section{Results}
\label{sec:results}

In this section we present accuracy and performance results of {\tt MadFlow} on
consumer and professional grade GPU and CPU devices. We focus on Leading Order
calculation for hadronic processes at $\sqrt{s}=13$ TeV. For this exercise we
select 5 processes with growing number of diagrams and channels in order to
emulate the behaviour in terms of complexity of a Next-to-Leading Order and
Next-to-Next to Leading Order computations. In particular, we consider: $gg
\rightarrow t \bar{t}$ (3 diagrams), $pp \rightarrow t \bar{t}$ (7 diagrams), $
p p \rightarrow t \bar{t} g $ (36 diagrams), $ p p \rightarrow t \bar{t} g g$
(267 diagrams) and $ p p \rightarrow t \bar{t} g g g$ (2604 diagrams).

Note that all results presented in this section have been computed using the
matrix elements generated by MG5\_aMC in double precision without any further
optimization, thus further releases of {\tt MadFlow} will address systematically
memory and performance optimization in order to achieve even better results.

All results presented in this section are obtained with {\tt madflow} 0.1, {\tt
vegasflow} 1.2.1, {\tt pdfflow} 1.2.1, MG5\_aMC 3.1.0, {\tt tensorflow} 2.5.0
for NVIDIA/Intel/EPYC systems with CUDA 11.3 drivers, and {\tt tensorflow-rocm}
2.4.1 with ROCm 4.2.0 drivers on Radeon/AMD systems.

\subsection{Accuracy}

\begin{figure}
    \centering
    \includegraphics[width=0.45\textwidth]{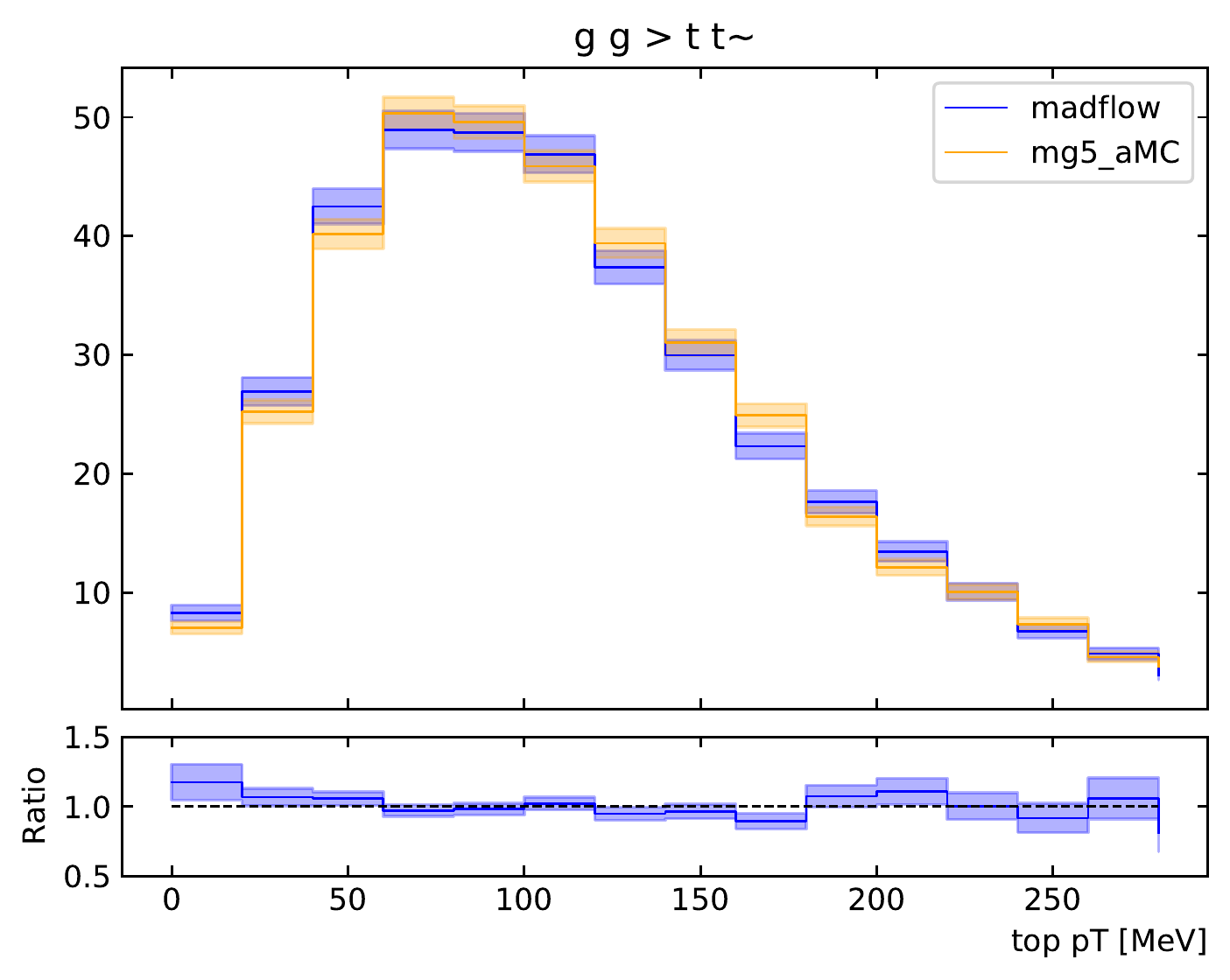}\\
    \includegraphics[width=0.45\textwidth]{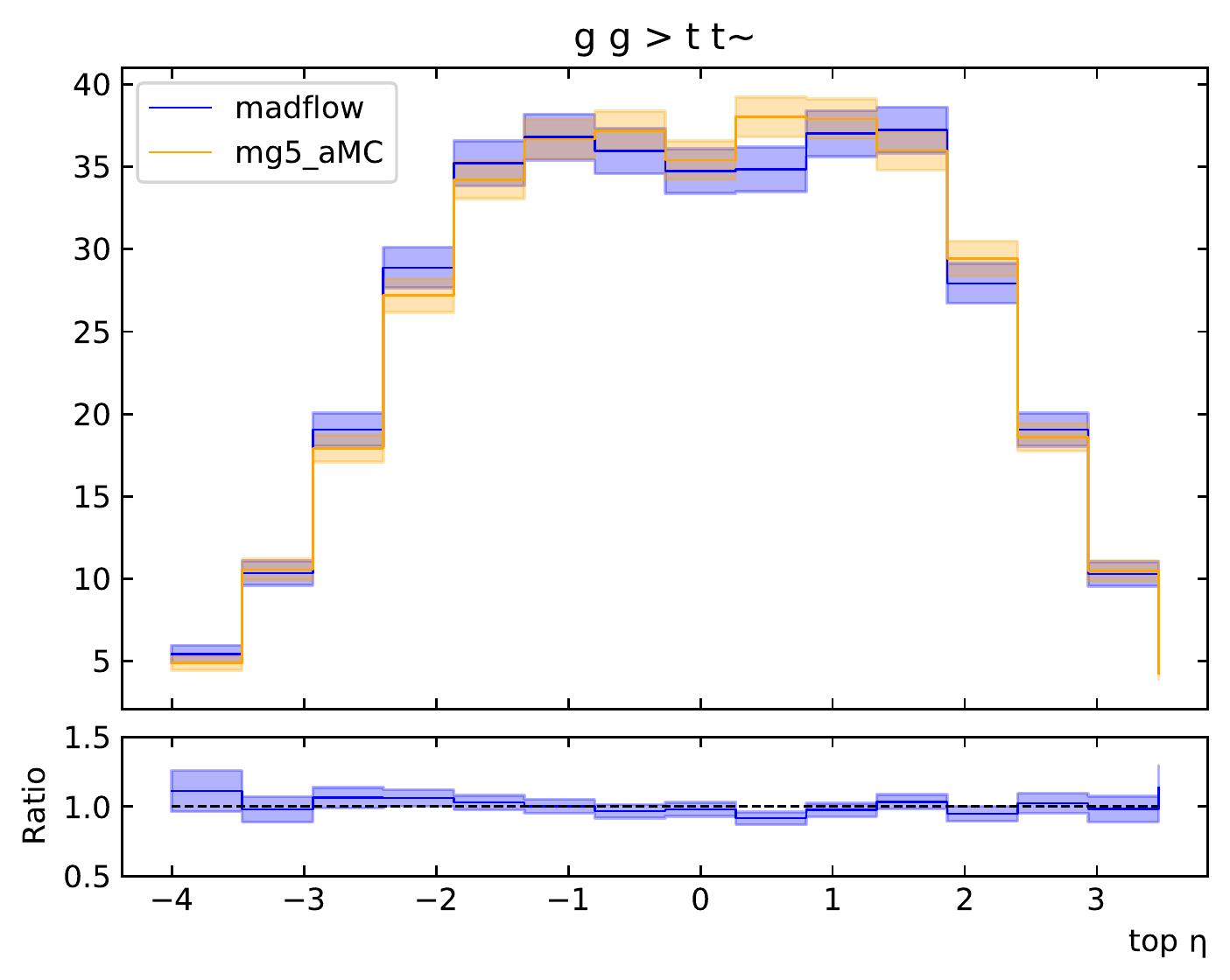}
    \caption{\label{fig:accuracy} Leading Order cross section differential on
        $p_{t,{\rm top}}$ (first row) and $\eta_{\rm top}$ (second row) for $gg
        \rightarrow t \bar{t}$ at $\sqrt{s}=13$ TeV. We compare predictions
        between {\tt MadFlow} (blue) and MG5\_aMC (orange). We observe that in
        both cases the distributions are in statistical agreement.}
\end{figure}

In Figure~\ref{fig:accuracy} we show an example of Leading Order cross section
differential on $p_{t, \rm top}$ and $\eta_{\rm top}$ for $gg \rightarrow
t\bar{t}$ at $\sqrt{s}=13$ TeV for predictions obtained with the original
MG5\_aMC integration procedure and the {\tt MadFlow} approach based on VegasFlow
and PDFFlow. In the first row we show the differential distribution in $p_{t,
\rm top}$ using the absolute scale in fb/GeV and the respective ratio between
both MC predictions, while in the second row we show the $\eta_{\rm top}$
distribution, confirming a good level of agreement between both implementations
for the same level of target accuracy between 2-5\% for each bin.

The results presented here are computed independently from each framework in
order to minimize communication bottlenecks between CPU-GPU. The plots are
constructed from unweighted events stored using the LHE approach described in
Section~\ref{sec:lhe}.

\subsection{Performance}

\begin{figure*}
    \centering
    \includegraphics[width=0.45\textwidth]{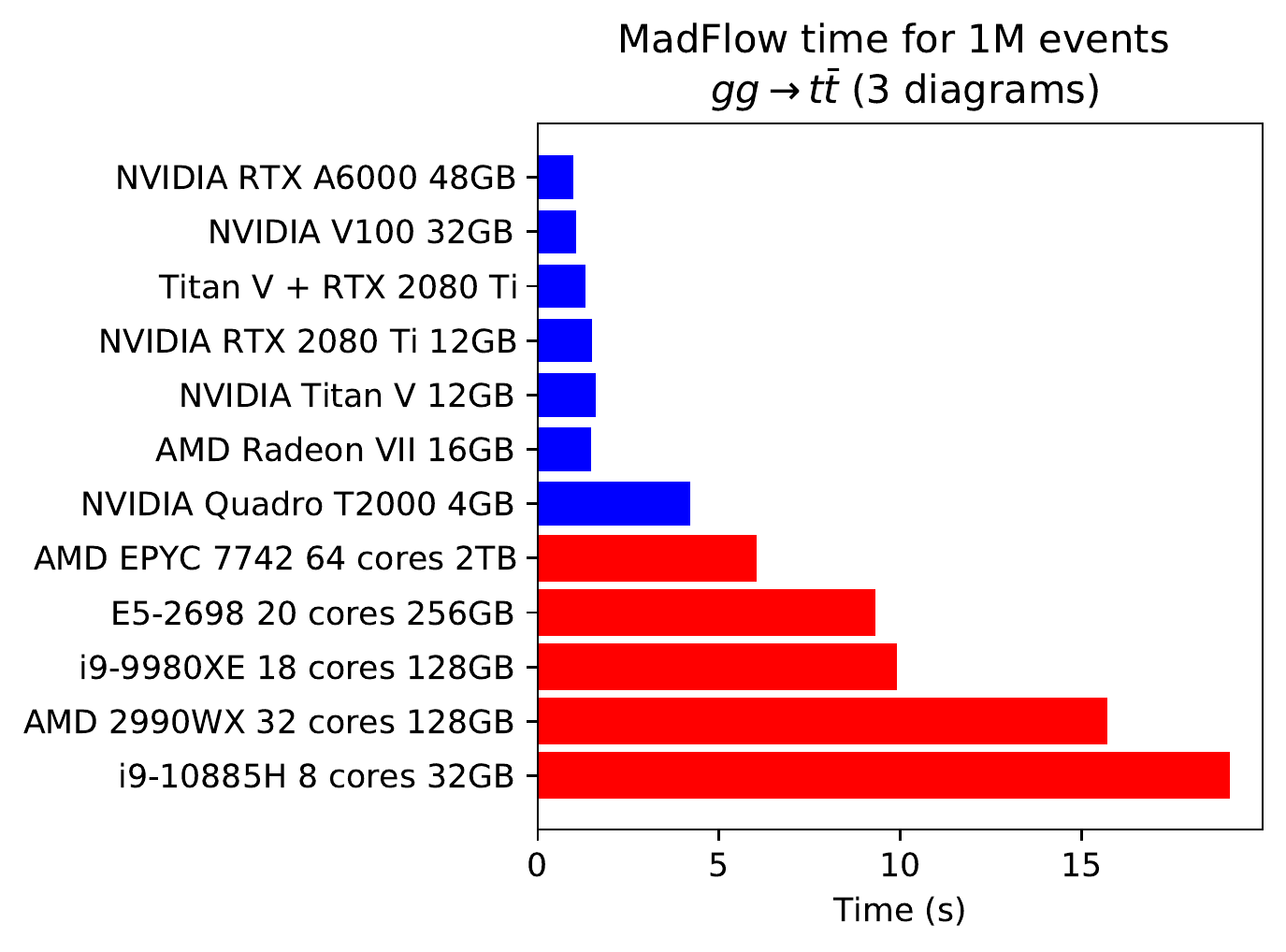}%
    \includegraphics[width=0.45\textwidth]{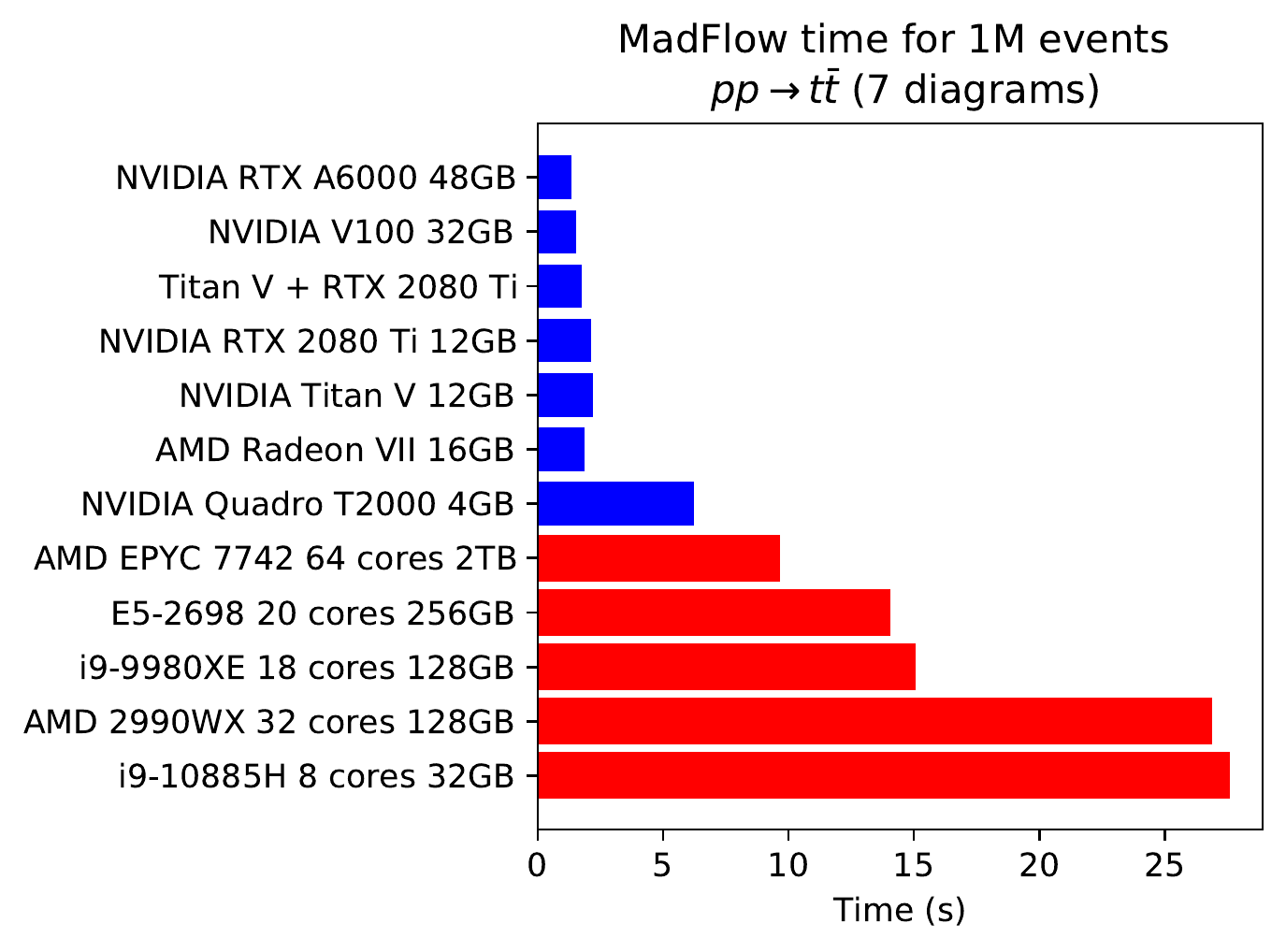}
    \includegraphics[width=0.45\textwidth]{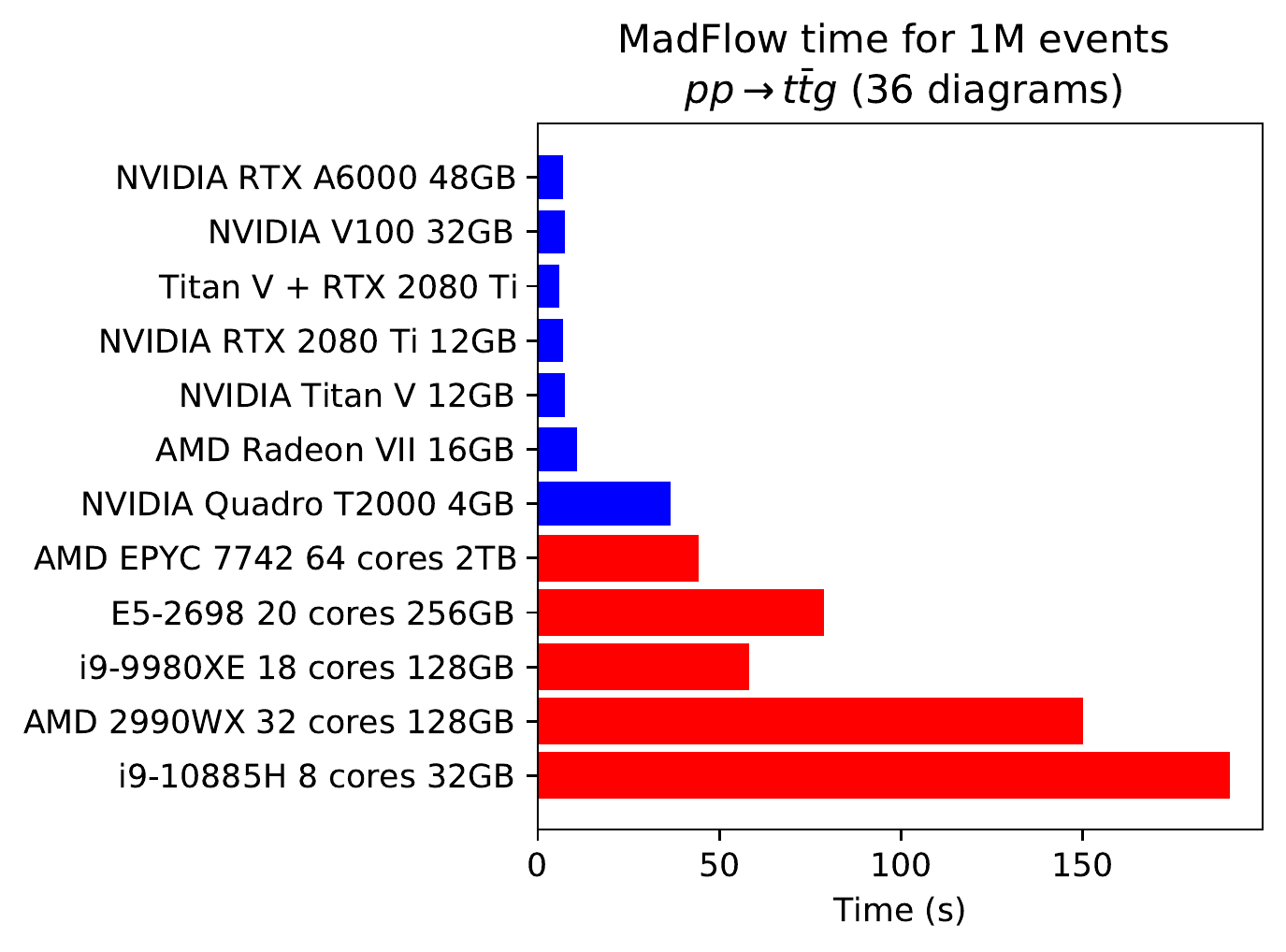}%
    \includegraphics[width=0.45\textwidth]{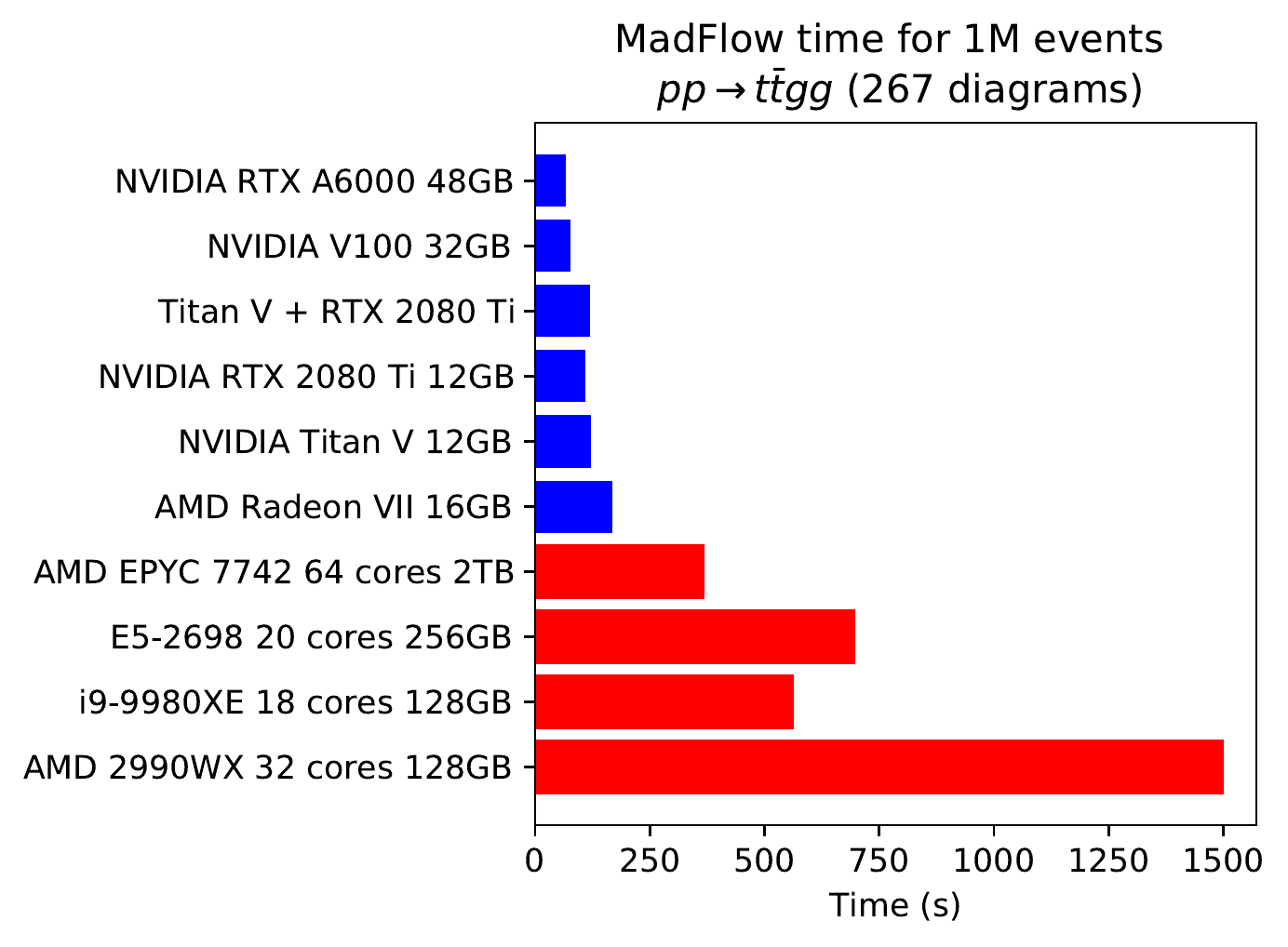}
    \caption{\label{fig:performance}Timings obtained with {\tt MadFlow} to
            evaluate events at Leading Order for $gg \rightarrow t\bar{t}$ (top
            left), $pp \rightarrow t \bar{t}$ (top right), $pp \rightarrow
            t\bar{t}g$ (bottom left) and $pp \rightarrow t\bar{t}gg$ (bottom
            right). We show results for consumer and professional grade GPUs
            (blue bars) and CPUs (red bars). For each device we quote the
            available RAM memory. We observe a systematic performance advantage
            for GPU devices.}
\end{figure*}

\begin{figure}
    \centering
    \includegraphics[width=0.45\textwidth]{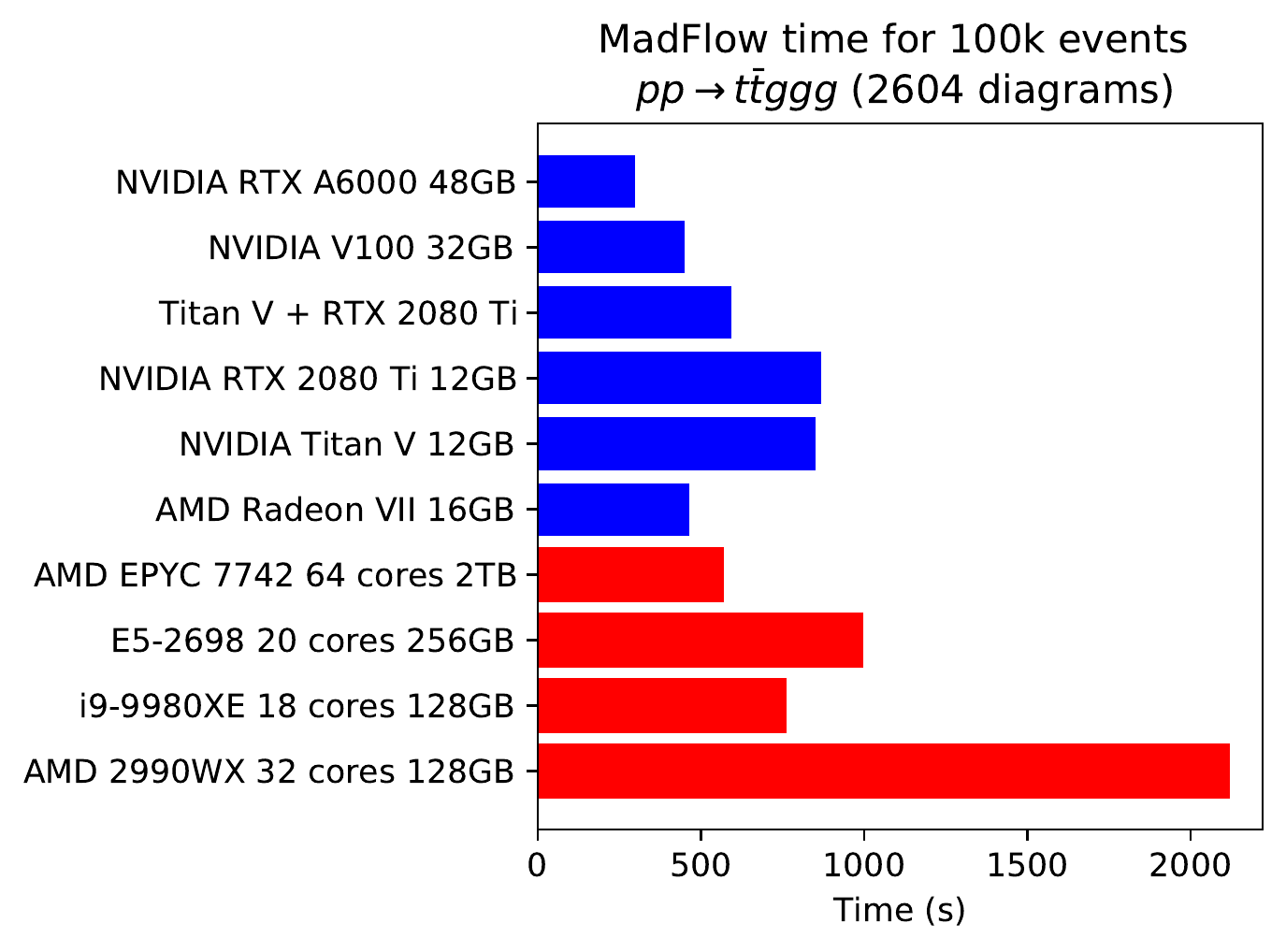}
    \caption{\label{fig:ttggg}Same as Figure~\ref{fig:performance} for $pp
            \rightarrow t\bar{t}ggg$ at Leading Order. We confirm that a large
            number of diagrams can be deployed on GPU and obtain relevant
            performance improvements when compared to CPU results.}
\end{figure}

In terms of performance, in particular evaluation time, in
Figure~\ref{fig:performance} we compare the total amount of time required by
{\tt MadFlow} for the computation of 1M events for the processes described
above: $gg \rightarrow t \bar{t}$ (3 diagrams), $pp \rightarrow t \bar{t}$ (7
diagrams), $ p p \rightarrow t \bar{t} g $ (36 diagrams), $ p p \rightarrow t
\bar{t} g g$ (267 diagrams). For all simulations, we apply a $p_t > 30$ GeV cut
for all out-going particles. We performed the simulation on multiple Intel and
AMD CPU configurations (red bars), together with NVIDIA and AMD GPUs (blue bars)
ranging from consumer to professional grade hardware. Blue bars show the
greatest performance of {\tt MadFlow} when running on GPU devices. We observe
that NVIDIA GPUs with the Ampere architecture, such as the RTX A6000,
out-perfoms the previous Tesla generation. We have observed that the performance
of the AMD Radeon VII is comparable to most professional grade GPUs presented in
the plot. The red bars show the timings for the same code evaluated on CPU using
all available cores. We confirm that GPU timings are quite competitive when
compared to CPU performance, however some top-level chips such as the AMD Epyc
7742, can get similar performance results when compared to general consumer
level GPUs, such as the Quadro T2000.
Note that in order to obtain good performance and going into production mode,
the {\tt MadFlow} user should adjust the maximum number of events per device, in
order to occupy the maximum amount of memory available.
We conclude that the {\tt MadFlow} implementation confirms a great performance
improvement when running on GPU hardware, providing an interesting trade-off in
terms of price cost and generated events.

In Figure~\ref{fig:ttggg} we present a preliminary example of simulation timings
for 100k events using {\tt MadFlow} as described above for $pp\rightarrow
t\bar{t}ggg$ with 2604 diagrams. The code generated for this example follows the
same procedure adopted for processes shown in Figure~\ref{fig:performance}. We
can remarkably confirm that {\tt MadFlow} results on GPU are competitive when
compared CPU results even for a such large number of diagrams (and thus required
GPU memory), taking into account that no custom code optimization has been
included. It is certainly true that the memory footprint and the overall performances of the code can (and should) be
 improved, e.g. by considering the Leading-Color approximation of the matrix element and/or
 possibly by performing a Monte-Carlo over color and helicity configurations, we believe that these
results confirm that GPU computation has a strong potential in HEP simulations at higher orders.

\subsection{Comparing to MG5\_aMC}

\begin{table}
    \centering
    \begin{tabular}{ l | c | c | c }
        Process & {\tt MadFlow} CPU & {\tt MadFlow} GPU & MG5\_aMC  \\ \hline
        $gg \rightarrow t\bar{t}$ & 9.86 $\mu$s & 1.56 $\mu$s & 20.21 $\mu$s \\
        $pp \rightarrow t\bar{t}$ & 14.99 $\mu$s & 2.20 $\mu$s & 45.74 $\mu$s \\
        $pp \rightarrow t\bar{t}g$ & 57.84 $\mu$s & 7.54 $\mu$s & 93.23 $\mu$s \\
        $pp\rightarrow t\bar{t}gg$ & 559.67 $\mu$s & 121.05 $\mu$s & 793.92 $\mu$s \\
        \hline
    \end{tabular}
    \caption{Comparison of event computation time for {\tt MadFlow}
        and MG5\_aMC, using an Intel i9-9980XE with 18 cores and 128GB of RAM
        for CPU simulation and the NVIDIA Titan V 12GB for GPU simulation.}
    \label{table:madflow_compare}
\end{table}

Finally, in Table~\ref{table:madflow_compare} we measure and compare the
required time per event for the processes discussed above using {\tt MadFlow}
and MG5\_aMC simulations on a Intel i9-9980XE CPU with 18 cores and 128GB of RAM
and a NVIDIA Titan V 12GB GPU.
As expected, we confirm that {\tt MadFlow} on GPU increases dramatically the
evaluated number of events per second.

Finally, as expected, the performance gain for GPUs when compared to CPU
decreases with the number of diagrams included in a given process thanks to the
amount of memory required to hold the computation workload. Such limitation
could be partially improved by using GPU models with larger memory, e.g. the new
NVIDIA A100 with 80GB, by compressing and optimizing the kernel codes before
execution~\cite{madgraph4gpu,Bothmann:2021nch}, and by using multi-GPU
configurations where portions of diagrams are distributed across devices.

\section{Outlook}
\label{sec:outlook}

In conclusion in this work we present {\tt MadFlow}, a new approach for the
generalization of Monte Carlo simulation on hardware accelerators. In
particular, the {\tt MadFlow} design provides a fast and maintainable code which
can quickly port complex analytic expressions into hardware specific languages
without complex operations involving several computer languages, tools and
compilers.
Furthermore, we confirm the algorithm effectiveness when running simulation on
hardware accelerators.

The {\tt MadFlow} code is open-source and public available on
GitHub\footnote{\url{https://github.com/N3PDF/madflow}}~\cite{juan_m_cruz_martinez_2021_4954376}.
The repository contains links to documentation for installation, hardware setup,
examples and development.

As an outlook, we plan to continue the development of {\tt MadFlow} as an
open-source library. Foreseen major improvements include: to replace the RAMBO
phase-space with more efficient solutions based on the process topology; to
investigate the possibility to accelerate integration using machine learning
techniques; finally, to set the stage for the the implementation of all required
changes to accommodate Next-to-Leading order computations.

\section*{Acknowledgements}

We thank the authors of ``Madgraph 4 GPU'', Olivier Mattelaer, Stefan Roiser and Andrea Valassi
for useful discussions. M.Z. wishes to thank Fabio Maltoni for discussions and suggestions.
S.C. and J.C.M are supported by the European Research Council under the European
Union's Horizon 2020 research and innovation Programme, grant agreement number
740006.
M.Z.~is supported by Programma per Giovani Ricercatori ``Rita Levi Montalcini''
granted by the Italian Ministero dell'Universit\`a e della Ricerca (MUR).
This project is supported by the UNIMI Linea 2A 2021 program.

\bibliographystyle{epj}
\bibliography{blbl.bib}

\end{document}